 \documentstyle[11pt,twoside,fleqn,espcrc1]{article}

\title{CHARMONIUM DECAY PHYSICS}

\author{Y.F. Gu\address{Institute of High Energy Physics, Beijing, China}
         and
        S.F. Tuan\address{University of Hawaii at Manoa, Honolulu, USA \\
\vspace{.10in}
           UH-511-944-99 \\
           BIHEP-EP1-99-04 \\
           October 1999}}       
\begin{document}
\maketitle

\begin{abstract}
Recent experimental results on the decays of charmonium, together with related 
physics issues, are reviewed. Some future prospects are described.
\end{abstract}

\section{INTRODUCTION}
The dramatic discovery of charmonium, the $J/\psi$ and its radial excitation
$\psi(2S)$, launched the modern era of particle physics. After a hiatus of
about one decade in the 1980's following a period of several-years of intense
experimental activity, charmonium physics has emerged again as one of the most
exciting areas of experimental high energy physics. A wealth of new data
in the last few years has changed greatly the face of this area.

As the ``hydrogen atom of strong interaction physics", charmonium states have
been studied in many experiments, which basically use three techniques:
formation and subsequent cascade decays from $e^{+}e^{-}$ annihilations, two
virtual photon interactions from high energy $e^{+}e^{-}$ collisions and
formation from $\bar{p}p$ collisions. At present the Beijing Spectrometer 
(BES) is the only experiment at the $e^{+}e^{-}$ collider (BEPC) to study 
charmonium physics around the $c\bar{c}$ threshold in $e^{+}e^{-}$ 
annihilations. The detectors at CESR and LEP, such as CLEO, DELPHI, L3, and 
OPAL, are performing experiments on two photon physics. The Fermilab 
experiment E760 and its upgraded experiment E835 are studying the direct 
formation of $c\bar{c}$ states in $\bar{p}p$ annihilations at the Fermilab 
Antiproton Accumulator Ring. Precision measurements of the $c\bar{c}$ system 
(masses, widths, decay rates, etc.) are important inputs to test the limit of 
PQCD and the order of magnitude of relativistic and radiative corrections.

Recent theoretical developments in effective field theories such as 
nonrelativistic QCD and heavy quark effective theory, lattice gauge theory, and
light front quantization suggest that it should be possible to place the theory
of charmonium on a rigorous foundation that is derived directly from QCD.

\section{REVIEW OF CHARMONIUM DATA}

In this section we will review the experimental data of charmonium states below
$D\bar{D}$ threshold. The subjects discussed are the mass, width, and other 
parameters of 1$^{3}S_{1}$, 2$^{3}S_{1}$, $^{3}P_{0,1,2}$, 1$^{1}S_{0}$, 
2$^{1}S_{0}$, and 1$^{1}P_{1}$ resonances. We emphasize on the results 
obtained since 1990.

\subsection{$1^3S_{1}$ : $J/\psi$}

A high precision measurement was performed by BES\cite{REF1} on leptonic 
branching fractions from a comparison of the exclusive and inclusive processes:
$\psi(2S) \rightarrow \pi^{+}\pi^{-}J/\psi$, with $J/\psi \rightarrow 
l^{+}l^{-}$ and $J/\psi \rightarrow anything$, which is luminosity independent
and almost free of QED backgrounds. The BES\cite{REF1} obtained values for
$B(J/\psi \rightarrow e^{+}e^{-}) = 5.90\pm0.05\pm0.10$\% and $B(J/\psi
\rightarrow \mu^{+}\mu^{-}) = 5.84\pm0.06\pm0.10$\%. Including BES data, the
new world average will have an error less than 1.5 \%, which is about a
factor of two improvement over the 1998 PDG value\cite{REF2}.

\subsection{$2^3S_{1}$ : $\psi(2S)$}

E760 has reported the first direct measurement of the total width of the 
$\psi(2S)$\cite{REF3}, $\Gamma = 306\pm36\pm16$ keV. Compared to the value
derived from a review of all previous data in 1992\cite{REF4}, $\Gamma =
243\pm43$ keV, the central value of E760 is larger. E760 performed new 
measurements on the branching fractions of $\psi(2S)$ decays to 
$J/\psi\pi^{+}\pi^{-}, J/\psi\pi^{0}\pi^{0},$ and $J/\psi\eta$ and claimed
that they are able to make measurements of $B(J/\psi\pi^{+}\pi^{-})$ and
$B(J/\psi\pi^{0}\pi^{0})$ with errors comparable to the world average
\cite{REF5}. However, as has been pointed out by Gu and Li\cite{REF6}, there
is logical inconsistency in handling of the computational procedure in Ref.
\cite{REF5}. As also pointed out by Gu and Li\cite{REF6}, the ratio of
$B(J/\psi\pi^{+}\pi^{-})/B(\mu^{+}\mu^{-})$ measured by E672/E706\cite{REF7}
as equal to $30.2\pm7.1\pm6.8$, was mistaken for 
$B(J/\psi\pi^{+}\pi^{-})/B(J/\psi\mu^{+}\mu^{-})$ in PDG 1998\cite{REF2}. They
thus suggested that we not use the 1998 PDG fit values of branching fractions
for the $\psi(2S)$ decays to $J/\psi$ plus anything\cite{REF6}. PDG will 
provide new fit values for $B(J/\psi + anything)$, $B(J/\psi + neutrals)$,
$B(J/\psi\pi^{+}\pi^{-})$, $B(J/\psi\pi^{0}\pi^{0})$, $B(J/\psi\eta)$,
$B(\gamma\chi_{c0})$, $B(\gamma\chi_{c1})$, and $B(\gamma\chi_{c2})$ in the 
next edition by removing the E760 data and correcting the above mistake
\cite{REF8}.

Using the world's largest data sample of $\psi(2S)$, BES has measured 
$\psi(2S)$ branching fractions for a large number of hadronic final states
- many for the first time[9-12]. 
The results for 2-body (light) meson final states will be discussed in
the next 
section in the context of hadronic decay puzzle.

\subsection{$^3P_{0,1,2}$ : $\chi_{c0}$, $\chi_{c1}$, $\chi_{c2}$}

The large sample of $\psi(2S)$ decays at BES permits the study of $\chi_{cJ}$
decays with unprecedented precision. Using many decay modes of the 
$\chi_{c0}$, BES has determined $M(\chi_{c0}) = 3414.1\pm0.6\pm0.8$ MeV
\cite{REF13}. The precision of this measurement represents a substantial
improvement over the existing PDG value of $3417.3\pm2.8$\cite{REF2}. BES 
also determined the $\chi_{c0}$ total width\cite{REF14}, $\Gamma(\chi_{c0})
= 14.3\pm2.0\pm3.0$ MeV, by selecting a $\pi^{+}\pi^{-}$ event sample, using
the precisely measured total width of the $\chi_{c2}$\cite{REF2} to determine
the detector resolution and a MC simulation to determine how the resolution
changes from $M(\chi_{c2})$ to $M(\chi_{c0})$. Compared with the only 
existing result of Crystal Ball\cite{REF15}, $\Gamma(\chi_{c0}) = 
13.5\pm3.3\pm4.2$ MeV, which is actually a combination of two measurements 
with large errors and of only marginal consistency (within $2.2\sigma$), the
uncertainty is now reduced from 40\% to 25\%.

$P$-wave charmonium states are directly accessible in $\bar{p}p$ 
annihilations. Precision measurements of the masses and the total widths of
the $\chi_{c1}$ and $\chi_{c2}$ resonances were performed by E760 using the
line shape method a few years ago\cite{REF16}. The results are given in 
Table~\ref{table:1}. These new values of the masses agree well with earlier 
measurements\cite{REF2}; the errors are reduced by more than a factor of two. 
The width of the $\chi_{c1}$ has been measured for the first time; the 
uncertainty on the $\chi_{c2}$ width has been reduced from about 40\% to about 
10\%.

\begin{table}[ht]
\caption{E760 measurements of $\chi_{c1}$ and $\chi_{c2}$ parameters}
\label{table:1}
\begin{tabular}{|c|c|c|} \hline
$^3P_J$ state      &   Mass (MeV)                  &   Width (MeV) \\ \hline
$\chi_{c1}$         &   3510.53$\pm 0.04 \pm 0.12$  &   0.88$\pm 0.11 \pm
0.08$ \\ \hline
$\chi_{c2}$         &   3556.12$\pm 0.07 \pm 0.12$  &   1.98$\pm 0.17 \pm
0.07$ \\ \hline
\end{tabular}
\end{table}
While there are only upper limits on $\gamma\gamma$ partial widths for the 
$\chi_{c0}$ resonance exist so far\cite{REF14,REF15}, there are a number 
of measurements for the $\chi_{c2}$ made by L3\cite{REF18,REF19}, 
E835\cite{REF20}, OPAL\cite{REF21}, CLEO\cite{REF2,REF22}, E760\cite{REF2}, 
and TPC\cite{REF2} since 1990. For $\chi_{c0}$, the limits are 
$\Gamma_{\gamma\gamma} < 6.2$ keV reported by CLEO and $ < 5.5$ keV 
reported recently by L3 (both 95\% C.L.); the only branching fraction
measured by Crystal Ball was never actually published\cite{REF23}.
The results for $\chi_{c2}$ are summarized in Table~\ref{table:2}. Among these
data, the central value of $\Gamma_{\gamma\gamma}$ for $\chi_{c2}$ differs 
significantly, and measurements at the $e^{+}e^{-}$ colliders (LEP,CESR,PEP)
seem all larger than the E760/E835 results obtained in $\bar{p}p$
annihilations. New measurements are still required. A reduction in the 
discrepancy found between the $\chi_{c2}$ data will be of fundamental 
importance to guide the extraction of theoretical parameters from the data.

\begin{table}[ht]
\caption{$\gamma \gamma$ partial width for $\chi_{c2}$}
\label{table:2}
\begin{tabular}{|l|c|} \hline
Experiment   &  $\Gamma_{\gamma\gamma} (\chi_{c2})$ (keV)  \\ \hline
L3[18]       & $<$ 1.4 (95\% C.L)    \\ \hline
E835 [20]    & 0.311$\pm$ 0.041$\pm$0.031 (prelim) \\ \hline
L3[19]       & 1.02 $\pm$0.40$\pm$0.15 $\pm$0.09  \\ \hline
OPAL[21]     & 1.76 $\pm$0.47$\pm$0.37$\pm$0.15   \\ \hline
CLEO[22]     & 0.7 $\pm$0.2$\pm$0.1$\pm$0.2   \\ \hline
CLEO[2]      & 1.08 $\pm$0.30$\pm$0.26   \\ \hline
E760[2]      & 0.321 $\pm$0.078$\pm$0.054   \\ \hline
TPC[2]       & 3.4 $\pm$1.7$\pm$0.9   \\ \hline
\end{tabular}
\end{table}
BES performed the first measurement of the branching fraction and the partial
width for $\chi_{c0} \rightarrow \bar{p}p$\cite{REF14}. After publication of
the BES results for all $\chi_{cJ} \rightarrow \bar{p}p$ branching fractions,
E835 also reported its first measurements on $\chi_{c0}$ and new results on
$\chi_{c1}$ and $\chi_{c2}$\cite{REF20}. The results are compared in 
Table~\ref{table:3}
with BES measurements. One notes that E760/E835 results for all 
$B(\chi_{cJ} \rightarrow \bar{p}p)$ are systematically higher (tantalizingly
large! - C. Quigg\cite{REF24}) with respect to BES results, though both have
large errors. It has been pointed out\cite{REF24} that in both experiments
the $B_{\bar{p}p}$ and $\Gamma_{\bar{p}p}$ are derived from the product of
branching fractions $B_{in} \times B_{out}$ and from $B(\psi(2S) \rightarrow
\gamma\chi_{cJ})$ and $B(\chi_{cJ} \rightarrow \gamma J/\psi)$ respectively.
New measurements of these two branching fractions would be desirable to
exclude one possible origin of such inconsistency.

\begin{table}[ht]
\caption{Comparison of BES and E760/E835 results for $\chi_{cJ}$  decays
B($\bar{p} p) \times 10^{4}$ (left) and $\Gamma (\bar{p} p)$ in
keV (right).}
\label{table:3}
\begin{tabular}{|c|c|c|c|c|} \hline
$^3P_J$  & BES  &   E760/E835   &   BES  &   E760/E835 \\ \hline
$\chi_{c0}$    &  1.59 $\pm$0.43$\pm$ 0.53    &   4.82$^{+0.97 +
2.08}_{-0.81-1.12}$   & 2.3$\pm$1.1    &   $8.0^{+1.9+3.5}_{-1.6-1.9}$ \\
\hline
$\chi_{c1}$     & 0.42$\pm$0.22$\pm$0.28  &   0.78$\pm$0.10$\pm$0.11    &  $0.037\pm 0.032$ 
      & $0.069\pm0.009$   \\
      &      &(E835)    &    & $\pm0.010$ (E835) \\
     &       & $0.86 \pm 0.12$ (E760)      &    &  $0.076  \pm 0.010$ \\
    &        &                           &    & $\pm 0.005$ (E760)  \\
\hline
$\chi_{c2}$   &   $0.58\pm$0.31$\pm$0.32   &  0.91$\pm$0.08$\pm$0.14  &
0.116 $\pm$0.090     &    $0.180\pm 0.016$  \\
&     &  (E835)      &     & $\pm 0.026$ (E835)  \\
   &    &  $1.00\pm 0.11$ (E760)     &    & $0.197\pm0.018$ \\
&    
  &     &    &         $\pm 0.016 (E760)$  \\ \hline
\end{tabular}
\end{table}

BES studied many other hadronic decays of $P$-wave charmonium states, and 
determined altogether more than 30 branching fractions for $\chi_{c0}$,
$\chi_{c1}$ and $\chi_{c2}$\cite{REF12,REF13} using the PDG values for
$B(\psi(2S) \rightarrow \gamma\chi_{cJ})$\cite{REF2}. Among them 15 were
measured for the first time.  

\subsection{$1^{1}S_{0}$: $\eta_{c}$}

In spite of a number of measurements on the mass of $\eta_{c}$, the value
remains ambiguous in PDG 1998 edition\cite{REF2}. The PDG average there is 
based on a fit to 7 measurements with poor internal consistency and the
confidence level is only 0.001. The measurement of E760\cite{REF25} disagrees
with the value of DM2\cite{REF26} by almost $4\sigma$ and is almost 10 MeV
different from 1994 PDG average. The change will cause a shift in the value
of the hyperfine splitting for the $S$-wave charmonium states which, in turn,
are important in understanding the spin-spin forces.

The value of $M(\eta_{c})$ determined recently by BES using several decay
modes of the $\eta_{c}$\cite{REF13} is in excellent agreement with DM2 data
and is $2.4\sigma$ below the E760 result. More recently, L3 has also reported
their measurement on the $\eta_{c}$ mass\cite{REF18}, which agrees well both
with BES and DM2. A comparison of recent results is shown in 
Table~\ref{table:4}.
\begin{table}[ht]
\caption{Recent data on $\eta_c$ mass}
\label{table:4}
\begin{tabular}{|l|l|} \hline
Experiment    &  Mass of $\eta_c$ (MeV) \\ \hline
L3[18]        &   2974 $\pm$ 18  \\ \hline
BES[13]        &   2975.8 $\pm$ 3.9 $\pm$ 1.2  \\ \hline
E760[25]       &   $2988.3^{+3.3}_{-3.1}$  \\ \hline
DM2[26]        &   2974.4$\pm$1.9   \\ \hline
MARK3 [27]      &  2969 $\pm$4
$\pm$ 4 \\ \hline
\end{tabular}
\end{table}

New measurements on $\eta_c$ width was made by E760\cite{REF25}, $\Gamma =
23.9^{+12.6}_{-7.1}$ MeV, and improved by E835 afterwards\cite{REF20} with
$\Gamma = 17.8^{+7.2}_{-6.9}$ MeV (preliminary). The results are still larger
than previous measurements\cite{REF2} where for instance the Crystal Ball 
value in 1986 is listed as $11.5\pm4.5$ MeV, and the errors remain large. The
$\gamma\gamma$ partial width of $\eta_{c}$ was measured by a number of 
experiments, including L3\cite{REF18}, E760/E835\cite{REF2,REF20},
ARGUS\cite{REF2} and CLEO\cite{REF2,REF22}. Unfortunately, the data 
are not of sufficient precision to differentiate 
between theories. 
\subsection{$2^1S_{0}$: $\eta_{c}(2S)$}

After observation of a candidate of this state by Crystal Ball experiment at 
$3594$ MeV\cite{REF28}, it has been searched for in E835\cite{REF20}, 
BES\cite{REF13}, DELPHI\cite{REF29}, and L3\cite{REF18} recently. No evidence 
is found in the mass region around $3594$ MeV by any of the subsequent 
experiments. This appears to challenge the theoretical analysis of Barnes, 
Browder, and Tuan \cite{REF30} based on the relationship that hadronic 
branching fractions of $\eta_{c}$ and $\eta_{c}(2S)$ to the same exclusive 
final state channel could be equal\cite{REF31} and a nonrelativistic quark 
potential model calculation\cite{REF32} 
that $\Gamma(\eta_{c}(2S) \rightarrow \gamma\gamma) = 3.7$ keV.
The L3\cite{REF18} upper limit on $\Gamma_{\gamma\gamma}(\eta_{c}(2S)) <
2.0$ keV (95\% C.L.) is not yet a severe constraint since the model 
calculation of this partial width might only be good to a factor of 2 to
3\cite{REF32}. Search
for $\eta_{c}(2S)$ in the two photon process at CESR, which has already
delivered more than $11 fb^{-1}$ of integrated luminosity to CLEO, remains a
valuable goal. We must caution however that though the important observation
that $\gamma\gamma$ widths are not strongly suppressed with radial excitation
in any of the $q\bar{q}$ systems considered\cite{REF30}, to date no radial
excitations have been identified in $\gamma\gamma$ collisions, so for the
present this width calculation should be taken as a theoretical estimate in a
regime in which theory has not been tested.

\subsection{$^1P_{1}$: $h_{c}(1P)$}

E760 announced the discovery of this state at $3526.14$ MeV\cite{REF33}. In a
subsequent search for hidden charm states in $\pi^--$ and $p-Li$ interactions,
E705 reported the observation of a $J/\psi\pi^{+}\pi^{-}$ signal at $3.836$
GeV (possible $^{3}D_{2}$ state) and a $J/\psi\pi^{0}$ enhancement at $3.527$
GeV (possibly the $^{1}P_{1}$ state)\cite{REF34}. However, E672/E706 has
questioned the strong structure at $3.836$ GeV\cite{REF7}. It was also 
questioned by Barnes, Browder, and Tuan\cite{REF35} whether E705 could have
`confirmed' E760's discovery of $^{1}P_{1}$ state. E835 will continue this
work and look further with more data in the near future.

\section{CHARMONIUM HADRONIC DECAY PUZZLE}

This celebrated ``$\rho-\pi$'' puzzle with dramatic suppression of $\psi(2S)
\rightarrow VP$\cite{REF36,REF37} and $VT$\cite{REF9}, but apparent non 
suppression of $\psi(2S) \rightarrow AP, VS$ as well as isospin violating 
modes $\omega\pi^{0}$, $\rho\eta^{0}$ (with branching ratios in accord with 
PQCD ``14\%'' rule), has been mostly summarized in \cite{REF38}. We note in 
particular that BES has concluded that $\psi(2S) \rightarrow \omega\pi^{0}$ 
is {\bf larger} than (strong) isospin conserving, SU(3)-allowed, 
$\psi(2S) \rightarrow \rho\pi$ decay while large isospin violations are seen 
between branching fractions for charged and neutral 
$\psi(2S) \rightarrow K^{*}\bar{K}$ decays.

The failure of most theoretical models up to 1990 have been summarized in 
Table~\ref{table:6}, while those proposed in recent years have been 
summarized in \cite{REF39}. Actually the model of Li-Bugg-Zou (LBZ for short)
\cite{REF40}, though fortuitous\cite{REF39}, cannot be ruled out. 
Based on final state interaction FSI, Suzuki\cite{REF41} nevertheless pointed 
out that their numerical computation picks two completely arbitrary 
intermediate states in estimating the FSI effects. One can in fact get 
almost any number by selecting intermediate states of one's choice. 
Unfortunately the intermediate states picked by LBZ are in fact heavily 
dominated by other intermediate states. Hence LBZ model does not answer the 
question of which specific FSI is responsible for the puzzle.
\begin{table}[htb]
\caption{Theoretical models up to 1990 and related experimental results.}
\label{table:6}
\begin{tabular}{|l|l|l|} \hline

Model    &  Predictions    & Experimental results   \\ \hline
Brodsky-Lepage-Tuan        & Hadron helicity conserved:  &  \\
(1987);   & -$\psi(2S) \rightarrow$ VP modes  & -$\psi(2S) \rightarrow \omega
\pi^0$  \\
Hou-Soni   & \ \ suppressed; &  \ \ not suppressed;  \\
(1983)     & -$\psi(2S)\rightarrow$ VT modes & -$\psi(2S) \rightarrow$ VT \\
 & \ not suppressed. & \ \ suppressed. \\
 & $J/\psi$-glueball mixing:        &  \\
 & -$J/\psi$ shape distorted;        &  -Not seen; limits set; \\
 & -Search in $\psi(2S)\rightarrow\pi^+ \pi^-$\sl{O} & -Not seen; limits set.\\
 & -$J/\psi\rightarrow\phi f_0$ enhanced.  & -$\psi(2S)\rightarrow\phi f_0$ \\
 &     &  \ \ not suppressed.   \\ \hline
Chaichian-Tornqvist    &   Energy dependent    &    \\
(1989)    &  exponetial form factor:     &  \\
&   -$\psi(2S)$ 2-body meson modes  &   -$b_1 \pi$  and $\phi f_0$  \\
   & \ suppressed;     &  \ \  not suppressed;  \\
   &  -B$(\psi (2S) \rightarrow \rho \pi ) = 7 \times 10^{-5}$  & -B$(\psi
(2S) \rightarrow \rho \pi)$  \\
&  &  $ \ < 2.8 \times 10^{-5}$. \\ \hline
Pinsky    & $\psi (2S) \rightarrow$ VP are hindered     &  \\
(1990)   & M1 transitions:   &  \\
  &   -B$(\psi (2S) \rightarrow \gamma \eta') = 9 \times 10^{-6}$;
  & -B$(\psi (2S) \rightarrow \gamma \eta')$  \\
    &   &  \ = 150  $\times 10^{-6}$;  \\
 &   -B$(\psi (2S) \rightarrow \rho \pi ) = 4 \times 10^{-5}$;  &
    -B$(\psi (2S) \rightarrow \rho \pi )$ \\
   &   & \ $< 2.8 \times 10^{-5}$;  \\
  & -$\psi(2S)\rightarrow \omega f_2$ & -$\psi(2S)\rightarrow \omega f_2$ \\
  &    \  not suppressed.   & \ \ suppressed.   \\ \hline
\end{tabular}
\end{table}    

The most recent model of G\'{e}rard and Weyers\cite{REF42} has the following
problems. (a) The BES data\cite{REF11} that for $\psi(2S) \rightarrow AP$ with
$K_{1}(1270)\bar{K}$ (large) and $K_{1}(1400)\bar{K}$ (small), cannot be 
clearly understood in the model. (b) The universality assumption for three-
gluon hadronization of $\psi(1S)$ remains doubtful\cite{REF43}. For instance
the three gluons from $\psi(1S)$ must certainly hadronizes to say $VP$ and
$VT$ final states in different ways. Can the phase really be the same? (c) The
model emphasizes on $\psi(2S) \rightarrow AP, AS$ final states to leading
order. Hence {\bf unsuppressed} $\psi(2S) \rightarrow \phi f_{0}(980)$
\cite{REF44,REF12} a $VS$ mode, and $\psi(2S) \rightarrow K^{*0}\bar{K}^{*0}$
a $VV$ mode\cite{REF12} would appear to be at variance with the model.
There is a need for further concerted effort on both theoretical and 
experimental side to provide a {\bf solution} to the $J/\psi/\psi(2S)
\rightarrow \rho\pi$ puzzle.

The $\rho-\pi$ puzzle motivated the important discovery of long-distance
(large phase) FSI physics\cite{REF45,REF46,REF47} from $J/\psi \rightarrow
VP, PP, B\bar{B}$ data. {\bf Its resolution remains very important}. For 
instance Suzuki noted\cite{REF45} that in $B$-meson decays knowledge of much
higher precision will be needed for FSI phases above the inelastic thresholds,
a nearly impossible task for theoretical computation/extraction from 
scattering data. {\bf Parameters of fundamental interactions} can then only
be extracted from data free from FSI (a severe limitation?). Also \cite{REF47}
stressed that FSI in nonleptonic $B$-decay has been an important unsolved
issue in direct search for $CP$ violations.

Rosner did {\bf significant} damage control\cite{REF48,REF49} for the future 
of $B$-Factory physics. He introduced (i) {\bf universal} FSI as consequence
that $\gamma$ and 3g amplitudes for $J/\psi$ are out of phase ($ \simeq 
\pi/2$) with each other. (ii) Connection is made with charmonium where strong
phase shifts in $B \rightarrow PP$ arise as result of strong absorptive 
effects in rescattering of $c\bar{c} \rightarrow$ light quarks\cite{REF48}.
(iii) Predicts (c.f. Table VI of \cite{REF48}) direct $CP$ asymmetries in
$B^{0}(\bar{B}^{0}) \rightarrow K^{+}\pi^{-}(K^{-}\pi^{+})$, $B^{\pm}
\rightarrow K^{\pm}\pi^{0}$ maximally $\sim$ 0.34. (iv) Emphasized decays of
neutral $B$ mesons to $CP$ eigenstates such as $J/\psi K^{0}_{S}$ and 
$\pi\pi$ can directly probe CKM phases, since their interpretation is
{\bf immune from strong FSI}. Hence recent measurement of $sin 2\beta$
\cite{REF50}, a $CP$ violating parameter, remains valid. (v) Suzuki
\cite{REF51} has continued this favorable ambiance with a very recent paper
on testing direct $CP$ violation of standard model {\bf without knowing strong
phases}.

Since {\bf large $CP$ asymmetry would require large FSI}\cite{REF43}, CLEO III
with a single ring and a well tried detector (suitably upgraded) could be
decisive in the study of $CP$ asymmetries for $B^{0}(\bar{B}^{0}) \rightarrow
K^{+}\pi^{-}(K^{-}\pi^{+})$, $B^{\pm} \rightarrow K^{\pm}\pi^{0}$ before
year end.

Emphasis on $J/\psi/\psi(2S)$ physics should not detract us from the
significant physics to be done in the open charm domain. For instance,
the $D$ can be fully reconstructed in 
$\psi^{\prime\prime}(3.772) \rightarrow D\bar{D}$, while 
$\bar{D} \rightarrow \mu+\nu$ can be deployed to measure $f_{D}$. 
Currently $f_{D} < 290$ MeV and $f_{D_{s}} = 250$MeV,
while $SU(3)$ breaking suggests $f_{D_{s}}/f_{D} = 1.1 - 1.25$, so $f_{D}
\sim 200 - 220$ MeV (an attainable experimental goal). Grinstein\cite{REF52}
says that up to 5\%, we have
\begin{equation}
f_{B_{s}}/f_B \simeq f_{D_{s}}/f_D, \ \mbox{and} \ [\Delta M_s/\Delta M_d]^{1/2}
\simeq
\frac{\mid V_{ts}\mid}{\mid V_{td}\mid} (f_{B_{s}}/f_B)
\end{equation}
where $\Delta M_{s}$ and $\Delta M_{d}$ are $B\bar{B}$ splittings in
strange/non strange $B$ respectively. So we are again back to 
{\bf fundamentals} of measuring CKM matrix elements! Finally with the advent 
of a Tau-Charm Factory, we must not forget about exploration of molecular
charmonium states as discussed recently\cite{REF53}.

\section{FUTURE PROSPECTS} 

E835 experiment will continue to take data during 1999-2000 period, with 
$20pb^{-1}$ accumulation of $\chi_{c0}$, $100pb^{-1}$ of $\eta_{c}(2S)$, and
$200pb^{-1}$ of $\psi(^{1}P_{1})$ anticipated respectively. At the forthcoming
run at BES, accumulation of $5\times10^{7} J/\psi$ are expected, while there
is a proposal for $2\times10^{7} \psi(2S)$ run. It is to be hoped that there
will also be a run at $\psi^{\prime\prime}(3.772)$ for open charm study. Then
there are the $B$-Factories Babar/PEP II, Belle/KEK-B {\bf and CLEO III}. Many
in the high energy physics community feel that charm spectroscopy both below
and above $D\bar{D}$ threshold is fascinating and badly needs a new high
statistics facility. A Tau-Charm Factory with luminosity about two orders of
magnitude higher than the BEPC would fill this need.



\begin{thebibliography}{99}
\bibitem{REF1} BES Collaboration, J.Z.~Bai {\em et al.}, Phys. Rev. {\bf D} 58
(1998) 092006.
\bibitem{REF2} Particle Data Group, C.~Caso {\em et al.}, Eur. Phys. J. {\bf
C} 3 (1998) 1.
\bibitem{REF3} E760 Collaboration, T.A.~Armstrong {\em et al.}, Phys. Rev.
{\bf D} 47 (1993) 772.
\bibitem{REF4} Particle Data Group, K.~Hikasa {\em et al.}, Phys. Rev. {\bf D}
45 (1992), 1 June, Part II.
\bibitem{REF5} E760 Collaboration, T.A.~Armstrong {\em et al.}, Phys. Rev.
{\bf D} 55 (1997) 1153.
\bibitem{REF6} Y.F.~Gu and X.H.~Li, Phys. Lett. {\bf B} 449 (1999) 361.
\bibitem{REF7} E672/E706 Collaboration, A.~Gribushin {\em et al.}, Phys. Rev.
{\bf D} 53 (1996) 4723.
\bibitem{REF8} C.~Amsler (PDG), private communication (1998).
\bibitem{REF9} BES Collaboration, J.Z.~Bai {\em et al.}, Phys. Rev. Lett.
{\bf 81} (1998) 5080.
\bibitem{REF10} BES Collaboration, J.Z.~Bai {\em et al.}, Phys. Rev. {\bf D}
58 (1998) 097101.
\bibitem{REF11} BES Collaboration, J.Z.~Bai {\em et al.}, Phys. Rev. Lett.
{\bf 83} (1999) 1918.
\bibitem{REF12} F.~Liu (BES Collaboration), these Proceedings.
\bibitem{REF13} BES Collaboration, J.Z.~Bai {\em et al.}, Phys. Rev. {\bf D}
60 (1999) 072001.
\bibitem{REF14} BES Collaboration, J.Z.~Bai {\em et al.}, Phys. Rev. Lett.
{\bf 81} (1998) 3091.
\bibitem{REF15} Crystal Ball Collaboration, J.~Gaiser {\em et al.}, Phys. Rev
{\bf D} 34 (1986) 711.
\bibitem{REF16} E760 Collaboration, T.A.~Armstrong {\em et al.}, Nucl. Phys.
{\bf B} 373 (1992) 35; Phys. Rev. Lett. {\bf 68} (1992) 1468.
\bibitem{REF17} CLEO Collaboration, W.-Y.~Chen {\em et al.}, Phys. Lett.
{\bf B} 243 (1990) 169.
\bibitem{REF18} L3 Collaboration, M.~Acciarri {\em et al.}, CERN-EP/99-072,
May 21, 1999, submitted to Phys. Lett. {\bf B}.
\bibitem{REF19} L3 Collaboration, M.~Acciarri {\em et al.}, Phys. Lett.
{\bf B} 453 (1999) 73.
\bibitem{REF20} N.~Pastrone (E835 Collaboration), Hadron Spectroscopy, 
Frascati, March 8-12, 1999.
\bibitem{REF21} OPAL Collaboration, K.~Ackerstaff {\em et al.}, Phys. Lett.
{\bf B} 439 (1998) 197.
\bibitem{REF22} V.~Savinov and R.~Fulton (CLEO Collaboration), Proc. X
Intern. Workshop on Photon-Photon Collisions, p. 203 (ed. D.J.~Miller
{\em et al.}, Sheffield, England, 1995).
\bibitem{REF23} R.~Lee (Crystal Ball Collaboration), Ph.D thesis, SLAC 282
(1985), unpublished.
\bibitem{REF24} C.~Quigg, Fermilab-conf-98/390-T(1998).
\bibitem{REF25} E760 Collaboration, T.A.~Armstrong {\em et al.}, Phys. Rev.
{\bf D} 52 (1995) 4839.
\bibitem{REF26} DM2 Collaboration, D.~Bisello {\em et al.}, Nucl. Phys.
{\bf B} 351 (1991) 1.
\bibitem{REF27} MARK3 Collaboration, Z.~Bai {\em et al.}, Phys. Rev. Lett. 
{\bf 65} (1990) 1309.
\bibitem{REF28} Crystal Ball Collaboration, C.~Edwards {\em et al.}, Phys.
Rev. Lett. {\bf 48} (1982) 70.
\bibitem{REF29} DELPHI Collaboration, P.~Abreu {\em et al.}, Phys. Lett.
{\bf B} 441 (1998) 479.
\bibitem{REF30} T.~Barnes, T.E.~Browder, and S.F.~Tuan, Phys. Lett. {\bf B}
385 (1996) 391.
\bibitem{REF31} K.T.~Chao, Y.F.~Gu, and S.F.~Tuan, Commun. Theor. Phys. 
{\bf 25} (1996) 471.
\bibitem{REF32} E.S.~Ackleh and T.~Barnes, Phys. Rev. {\bf D} 45 (1992) 232;
C.R.~M\"{u}nz, Bonn University Report TK-96-01 obtains/summarizes other
predictions between a factor of two to three smaller than Ackleh/Barnes.
\bibitem{REF33} E760 Collaboration, T.A.~Armstrong {\em et al.}, Phys. Rev.
Lett. {\bf 69} (1992) 2337.
\bibitem{REF34} E705 Collaboration, L.~Antoniazzi {\em et al.}, Phys. Rev.
{\bf D} 50 (1994) 4258.
\bibitem{REF35} T.~Barnes, T.E.~Browder, and S.F.~Tuan, UH-511-868-97 (1997).
\bibitem{REF36} M.E.B.~Franklin {\em et al.}, Phys. Rev. Lett. {\bf 51}
(1983) 963.
\bibitem{REF37} BES Collaboration, J.Z.~Bai {\em et al.}, Phys. Rev. {\bf D}
54 (1996) 1221.
\bibitem{REF38} F.A.~Harris (BES Collaboration), hep-ex/9903036 v2 9 Jul 1999; 
K.T.~Chao, Lepton-Photon Symposium '95, p. 106 [World Scientific Publishing 
(1996)].
\bibitem{REF39} S.F.~Tuan, hep-ph/9903332 v3 24 Mar 1999, Commun. Theor. Phys.
(in press); Y.F.~Gu and S.F.~Tuan, Mod. Phys. Lett. {\bf A} 10 (1995) 615.
\bibitem{REF40} X.-Q.~Li, D.V.~Bugg, and B.-S.~Zou, Phys. Rev. {\bf D} 55
(1997) 1421.
\bibitem{REF41} M.~Suzuki, private communication.
\bibitem{REF42} J.-M.~G\'{e}rard and J.~Weyers, hep-ph/9906357 14 Jun 1999.
\bibitem{REF43} J.L.~Rosner, private communication.
\bibitem{REF44} Y.F.~Gu (BES Collaboration), Proc. of DPF '96 (Minneapolis), 
Vol. 2, p.986 [World Scientific Publishing (1998)].
\bibitem{REF45} M.~Suzuki, Phys. Rev. {\bf D} 57 (1998) 5717.
\bibitem{REF46} G.~L\'{o}pez Castro {\em et al.}, hep-ph/9902300 10 Feb 1999;
AIP Conference Proceedings {\bf 342} (1995) 441.
\bibitem{REF47} M.~Suzuki, Phys. Rev. {\bf D} 60 (1999) 051501.
\bibitem{REF48} J.L.~Rosner, hep-ph/9903543 31 Mar 1999, Phys. Rev. {\bf D}
(in press).
\bibitem{REF49} J.L.~Rosner, hep-ph/9905366 17 May 1999.
\bibitem{REF50} G.~Bauer (representing the CDF Collaboration), hep-ex/9908055
19 Aug 1999.
\bibitem{REF51} M.~Suzuki, hep-ph/9908420 19 Aug 1999.
\bibitem{REF52} B.~Grinstein, Phys. Rev. Lett. {\bf 71} (1993) 3067; C.~Glenn
Boyd and B.~Grinstein, Nucl. Phys. {\bf B} 442 (1995) 205, {\em ibid} Nucl.
Phys. {\bf B} 451 (1995) 177.
\bibitem{REF53} S.F.~Tuan, hep-ph/9903342 12 Mar 1999, submitted to Phys. 
Lett. {\bf B}.

\end{thebibliography}
\end{document}